\documentclass[aps,prd,a4paper,twocolumn,showpacs,showkeys,preprintnumbers,amsmath,amssymb]{revtex4}
\usepackage{graphicx}
\usepackage{bm}
\usepackage{dcolumn}
\usepackage{color}
\usepackage{amsmath}

\begin{document}


\title{Hausdorff dimension of a particle path in a quantum manifold}


\author{Piero Nicolini}
\email[]{nicolini@th.physik.uni-frankfurt.de}

\author{Benjamin Niedner\footnote{Current address: Theoretical Physics Group, Imperial College London
Prince Consort Road, London SW7 2AZ, UK}}
\email[]{bjn10@imperial.ac.uk
}

\affiliation{Frankfurt Institute for Advanced Studies (FIAS), \\Institut f\"{u}r Theoretische Physik, Johann Wolfgang Goethe-Universit\"{a}t,\\Ruth-Moufang-Stra\ss e 1, 60438 Frankfurt am Main, Germany}

\date{\today}

\begin{abstract}
\noindent
After recalling the concept of the Hausdorff dimension, we  study the fractal properties of a quantum particle path. As a novelty we consider the possibility for the space where the particle propagates to be endowed with a quantum-gravity-induced minimal length. We show that the Hausdorff dimension accounts for both the quantum mechanics uncertainty and manifold fluctuations. In addition the presence of a minimal length breaks the 
self-similarity property of the erratic path of the quantum particle. Finally we establish a universal property of the Hausdorff dimension as well as the spectral dimension: They both depend on the amount of resolution loss which affects both the path and the manifold when quantum gravity fluctuations occur.
\end{abstract}


\maketitle


\noindent
Fractals are often employed to describe the nature of a quantum manifold. Indeed one of the widely expected features in quantum gravity is the appearance of spacetime fluctuations as far as distances comparable with the Planck length are probed.
Fluctuations of this kind imply a loss of resolution: Distances smaller than the Planck length cannot be resolved, and one often speaks of minimal length effects when the spacetime passes from a low energy differential manifold to its Planck energy quantum configuration. Fractals nicely encode the idea of quantum fluctuations and loss of resolution.  Another attractive property of fractals is self-similarity, namely, the property of being exactly similar to a part of itself. This feature is connected to the concept of scale invariance, which seems to be supported by recent nonperturbative string theory developments like AdS/CFT and M theory.
To investigate the properties of a quantum spacetime one can employ technical tools which belong to the theory of fractals. As an example an important issue is the calculation of the spectral dimension, i.e., the manifold dimension perceived by a diffusion process. The way one formulates the presence of quantum fluctuations of the manifold is crucial and gives rise to a variety of expressions for the spectral dimension \cite{Carlip:2009kf,Ambjorn:2005db,Modesto:2008jz,Modesto:2009kq,Caravelli:2009gk,Magliaro:2009if,Lauscher:2005qz,
Dario,Nicolini:2010bj,Carlip:2010km}. As soon as the diffusion starts, small length scales of the manifold are probed and strong fluctuations emerge. The diffusion process is therefore subjected to a loss of resolution and the spectral dimension turns out to be smaller than the actual topological dimension of the manifold. In the case of a four-dimensional manifold, one of the crucial features is that at the Planck length the spectral dimension equals two, supporting the idea of a renormalizable character of the gravitational interaction as recently shown in  \cite{Modesto:2009qc}.

Another measure of the fractal nature of a manifold is provided by the Hausdorff dimension. One of the essential features of a fractal is that its Hausdorff dimension strictly exceeds its topological dimension \cite{falconer}.
As an example a quantum particle proceeds along an erratic path whose Hausdorff dimension is two, i.e., exceeding the dimension of a classical trajectory \cite{Abbott:1979bh}. For the stringy analogue, it has been shown that the world sheet  makes a transition to an excited configuration as far as length scales of order $(\alpha^\prime)^{1/2}$ are concerned. In this case the world sheet becomes a fractal surface of dimension three, since the energy in the excited state lets the string explore an additional dimension \cite{Ansoldi:1997cw,Ansoldi:1998ys,Aurilia:2002aw} (for further applications see \cite{ADJ}).
 While the spectral dimension accounts for the fractal character of the space where the diffusion takes place, the Hausdorff dimension for a quantum particle is just an indicator of the amount of uncertainty of a quantum path. Nothing is said about the intrinsic uncertainty, which any model of quantum spacetime should be endowed with
. In other words in \cite{Abbott:1979bh} the background space where the particle propagates is still a classical manifold. In this paper we want to do a step forward, by implementing the presence of a minimal length in the background space where a quantum particle propagates.
Therefore this paper has three main goals:
\begin{enumerate}
\item to provide an example where the Hausdorff dimension can be employed as an indicator of the amount of fluctuations of the manifold rather than of the particle path;
\item to disclose further properties of our method of implementing an effective minimal length in the manifold other than those we found by studying the spectral dimension in \cite{Modesto:2009qc};
\item to understand whether universal properties exist as a result of the study of both indicators of quantum fluctuations, i.e., the spectral dimension and the Hausdorff dimension.
\end{enumerate}

 \begin{figure}
  \begin{center}
  \includegraphics[height=6.5cm]{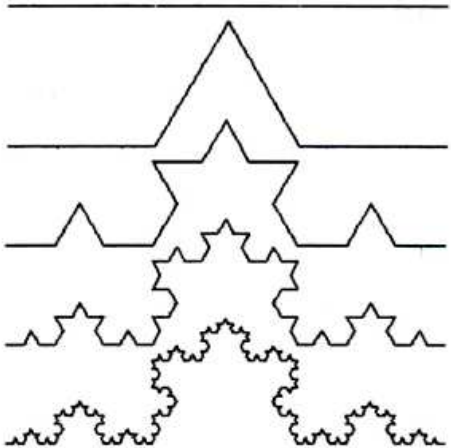}
  \hspace{0.2cm}
       \caption{\label{Plot1}Construction of the Koch curve. At each step, the middle third of each interval is replaced by the other two sides of an equilateral triangle.
  }
  \end{center}
   \end{figure}

Given this background we briefly recall the definition of the Hausdorff dimension. We start by considering the Koch curve in Fig. \ref{Plot1}. It is an example of an everywhere continuous but nowhere differentiable curve. We can construct the Koch curve as a final product of an infinite sequence of steps. At each step, the middle third of each interval is replaced by the other two sides of an equilateral triangle. As a result at each step the length of the curve increases by a factor $4/3$, so the final curve is infinitely long. However if we assume viewing the curve with a finite resolution $\Delta x$, many wiggles smaller than $\Delta x$ are neglected. Here the resolution $\Delta x$ has only a mathematical meaning related to the diameter of spheres covering the curve \cite{haus}.  As a result the observed length of the curve turns out to be finite, becoming infinite only in the limit $\Delta x\to 0$. Let $l$ be the length of the curve when the resolution is $\Delta x\neq 0$. Improving the resolution so that $\Delta x^\prime=(1/3)\Delta x$, we proceed along the next step of the curve and new wiggles become visible. As a consequence we will measure a new length $l^\prime=(4/3)l$. The length of the curve depends on the resolution at which the curve is examined. Therefore we cannot uniquely define the length of the curve in this way. To solve this problem Hausdorff proposed a new definition of length given by
\begin{equation}
L_H=l(\Delta x)^{D_H-1}.
\end{equation}
Here $l$ is the usual length when the resolution is $\Delta x$, and $D_H$ is a real number chosen so that $L_H$ will be independent of $\Delta x$, at least in the limit $\Delta x\to 0$. The parameter $D_H$  is called the \textit{Hausdorff dimension}. When the Hausdorff length $L_H$ coincides with the usual length $l$, the Hausdorff dimension equals the topological dimension $d_{\mathrm {top}}=1$ of the curve.
For the Koch curve we can calculate the Hausdorff dimension by requiring $L_H=L_H^\prime$ namely,
\begin{equation}
 l^\prime(\Delta x^\prime)^{D_H-1}=l(\Delta x)^{D_H-1}.
\end{equation}
This implies that $D_H=\ln 4/\ln 3$. The fact that $D_H\neq 1$ identifies the curve as a fractal.

\begin{figure*}
 \includegraphics[width=13cm]{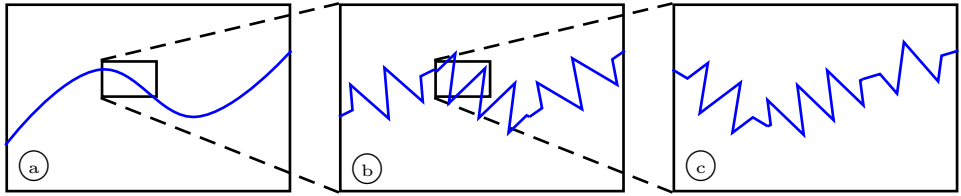}
  \hspace{0.2cm}
  		\caption{\label{magnification}Schematic view of the geometrical structure of the particle path. In (a) we have the classical regime and in (b) the quantum-mechanical regime. Upon further magnification in (c), the path exhibits the same structure.}
  \end{figure*}

We now switch from mathematics to physics. To do this we need to connect our ``mathematical'' resolution $\Delta x$ with a physically meaningful quantity. Along the lines of \cite{Abbott:1979bh} we consider the natural case of quantum mechanics. Typical paths of a quantum-mechanical particle are highly irregular on a fine scale. According to the Heisenberg uncertainty principle the more precisely the particle is located in space, the more its path will become increasingly erratic. If the localization of the particle is within a region of size $\Delta x$, an uncertainty will affect the momentum of order $\hbar/\Delta x$. In the language of fractals, this is equivalent to saying that paths for a quantum-mechanical particle are not those which admit a definite slope (velocity and therefore momentum) everywhere. For this reason in quantum mechanics we cannot properly speak of a particle path unless in the statistical sense. Suppose now to measure the position of a quantum particle at a sequence of times $t_0$, $t_1=t_0+\Delta t$, ..., $t_N=t_0+N\Delta t$, with $T=t_N-t_0=N\Delta t$. Then the length of the path will be
\begin{equation}
\left\langle l\right\rangle=N \left\langle \Delta l\right\rangle
\label{length}
\end{equation}
where
\begin{equation}
\langle \Delta l \rangle = \langle \psi|\hat{U}^{\dagger}(\Delta t)|\hat{\textbf{x}} | \hat{U}(\Delta t)| \psi \rangle
\label{Deltat}
\end{equation}
is the average distance which the particle travels in a time $\Delta t$, with $\hat{U}(t)=\exp(-i\hat{\textbf{p}}^2 t/2m\hbar)$ the free particle time evolution operator. Here the wave function of the particle
$\psi(\textbf{x})=\langle \textbf{x} | \psi \rangle$ takes into account the fact that a position measurement only localizes the particle within a region of size $\Delta x$. 
For later convenience we introduce the following dimensionless quantities: $\textbf{y}\equiv\textbf{x}/\Delta x$ and $\textbf{k}\equiv\textbf{p}\Delta x/\hbar$.
If we consider the case where the average momentum of the particle is zero, we obtain
\begin{eqnarray}
&&\langle \Delta l \rangle =\int \ d^d x|\textbf{x}|\left|\psi(\textbf{x}, \Delta t)\right|^2 \\
&&=(\Delta x) \int  d^dy |\textbf{y}| \left|\int \frac{d^d k}{(2\pi)^{d/2}}\langle \textbf{k} | \psi \rangle e^{\left( i\textbf{ky}-\frac{i\hbar\Delta t}{2m(\Delta x)^2}\textbf{k}^2\right)}
 \right|^2 . \nonumber
\end{eqnarray}
The actual profile of $\psi(\textbf{k})\equiv \langle \textbf{k} | \psi \rangle$ will not be important for our discussion. We just need a square integrable function, which can localize the particle to a region of size $\Delta x$. Thus a choice for $\psi(\textbf{k})$ is
\begin{equation}
\langle \textbf{k}|\psi\rangle=\left(\frac{2}{\pi}\right)^{d/4}e^{-\textbf{k}^2}.
\label{kprofile}
\end{equation}
Note that any other square integrable function can be approximated arbitrarily well by a linear combination of such Gaussian functions. As a result, one finds
\begin{eqnarray}
\langle \Delta l \rangle \propto \frac{\hbar\Delta t}{m\Delta x}\ \sqrt{1+\left(\frac{2m(\Delta x)^2}{\hbar \Delta t}\right)^2}.
\label{deltalc}
\end{eqnarray}
Recalling that $\Delta x$ is our resolution parameter, we can vary $\Delta x$ by keeping $\Delta t$ fixed. In the case $\Delta x\to 0$, (\ref{deltalc}) reduces to 
\begin{eqnarray}
\langle \Delta l \rangle \propto \hbar\Delta t/m\Delta x,
\end{eqnarray}
a result supported by the uncertainty principle. As a consequence
\begin{equation}
\left\langle l\right\rangle\propto \hbar T/m\Delta x,
\label{length2}
\end{equation}
a length which is ill defined since it depends on the detection resolution $\Delta x$ and diverges in the limit $\Delta x\to 0$.  This is a sign which confirms the fractal character of the path. Along Hausdorff's lines we can define a new length as
\begin{equation}
\langle  L_H \rangle =\left\langle l\right\rangle(\Delta x)^{D_H-1}
\label{hauslength}
\end{equation}
which turns out to be independent of $\Delta x$ if $D_H=2$. The path of a quantum particle is therefore a fractal of dimension two. We can check the other interesting property of fractals: self-similarity.  In the case of the Koch curve self-similarity is evident by increasing the resolution $\Delta x^\prime=(1/3)\Delta x$. In analogy the path of a quantum particle is self-similar if $\langle \Delta l \rangle \propto \Delta x$. This relation implies
\begin{equation}
\Delta t\propto m(\Delta x)^2/\hbar
\label{Deltatx}
\end{equation}
which naturally arises in the derivation of $\langle \Delta l \rangle$ as a consequence of the uncertainty principle in the energy-momentum relation $E=p^2/2m$.
The whole description of the path of a quantum particle can be generalized to the case when the particle has some nonzero average momentum $\textbf{p}_{\mathrm{av}}$. In this case the Hausdorff dimension is $D_H=1$ when the distances being resolved are much larger than the particle's wavelength, i.e., $\Delta x\gg \hbar/|\textbf{p}_{\mathrm{av}}|$. Conversely it is $D_H=2$ when the distances being resolved are much smaller than the particle's wavelength $\Delta x\ll \hbar/|\textbf{p}_{\mathrm{av}}|$. In the region between these limits the Hausdorff dimension $D_H$ is not well defined, since there is a ``phase transition'' from the classical to the quantum-mechanical path (see Fig. \ref{magnification}).

We are now ready to switch from quantum mechanics to quantum gravity. One of the most important features of quantum gravity is the appearance of an additional kind of uncertainty which prevents one from measuring positions to better accuracies than the Planck length. Indeed the momentum and the energy required to make such a measurement will itself modify the spacetime geometry at these scales \cite{DW}. This long held idea has been corroborated by the noncommutative character of open string end points on D-branes \cite{Seiberg:1999vs}. The specific feature of the presence of a minimal length in a spacetime manifold can be taken into account by means of effective theories too. Though they are not the full theory of quantum gravity, these effective formulations are particularly useful for getting reliable phenomenological scenarios in specific physical contexts \cite{NCApplications,kober}.  For instance, an effective minimal length has efficiently been included in the physics of evaporating black holes, by smearing out the curvature singularity at the origin and regularizing the terminal phase of the Hawking emission \cite{NCBHs} (for reviews see \cite{review1,review2}, and the references therein).
At the basis of these approaches there is the possibility of providing a delocalization of point like objects by the action of a nonlocal operator $e^{\ell^2\Delta_x}$ \cite{nonlocal}. Here $\ell$ is the minimal length and $\Delta_x$ is the Laplacian operator acting on a $d$-dimensional Euclidean manifold. As an example we consider the Dirac delta $\delta(\textbf{x})$ as a standard distribution for a point like object. By applying the nonlocal operator one finds
\begin{equation}
\delta(\textbf{x})\rightarrow e^{\ell^2\Delta_x}\delta(\textbf{x})=\rho_\ell(\textbf{x})
\label{smearing}
\end{equation}
where $\rho_\ell$ is the modified distribution due to the presence of the minimal length $\ell$. It turns out that the modified distribution is
\begin{equation}
\rho_\ell(\textbf{x})=\frac{1}{(4\pi \ell^2)^{d/2}}\ e^{-x^2/4\ell^2}
\end{equation}
which is nothing but a Gaussian distribution whose width equals $\ell$. Indeed this is the most narrow distribution which is admissible on a manifold endowed with a minimal length. In \cite{Modesto:2009qc} it has been shown that the primary effect of the presence of a minimal length in the dynamics of a diffusion process lies in a smearing of point like initial conditions as in (\ref{smearing}). We recall that a diffusion process is nothing but a Wick rotated quantum-mechanical probability evolution. It is therefore natural to extend the method we employed in \cite{Modesto:2009qc} for the spectral dimension to the case of the Hausdorff dimension too.
In the latter case, the action of the nonlocal operator $e^{(\ell^2/\hbar^2) \Delta_x}$ determines a modification of the integration measure of the momentum space representation of the wave function
\begin{equation}
\langle \textbf{x} | \psi \rangle = \int dV_\ell(p)\ \langle \textbf{p} | \psi \rangle e^{\frac{i}{\hbar}\textbf{p}\textbf{x}}
\end{equation}
where
\begin{equation}
dV_\ell(p)=\frac{d^d p}{(2\pi\hbar)^\frac{d}{2}} \ e^{-\frac{\ell^2}{\hbar^2} \textbf{p}^2}
\label{lvol}
\end{equation}
turns out to be squeezed for large momenta only, i.e., $|\textbf{p}|\gtrsim \hbar/\ell$. We stress that this kind of deformation of the integration measure is in agreement with what one requires in a generic effective approach to quantum gravity. Indeed the ultraviolet convergence is obtained in general as
\begin{equation}
dV(p)=\frac{1}{(2\pi\hbar)^\frac{d}{2}}\frac{d^d p}{\left(1+f( \textbf{p} ^2)\right)}
\end{equation}
where $f(\textbf{p}^2)$ depends on positive powers of the argument \cite{kober,KMM}.  The specific choice 
$f(\textbf{p}^2)=\beta {  \textbf{p}}^2$ corresponds to the case of the generalized uncertainty principle. The deformation we are following in (\ref{lvol}) is simply equivalent to $f(\textbf{p}^2)= e^{\frac{\ell^2}{\hbar^2} \textbf{ p}^2/2}-1$. This choice has two important virtues: It keeps the general features of ultraviolet convergence required by any effective approach to quantum gravity; it turns out to be the strongest possible suppression of higher momenta. 

We now have all the ingredients to investigate the fractal properties of the path of a quantum particle which propagates in a $d$-dimensional manifold endowed with a minimal length $\ell$. We start from the case where the average momentum of the particle is zero. The path will be affected by both the quantum mechanics uncertainty encoded in $\hbar$ and the quantum gravity uncertainty encoded in $\ell$. As a result the mathematical resolution $\Delta x$ will be related to both $\hbar$ and $\ell$. We want to understand the response of the manifold when it is probed at various regimes of energy, irrespective of the nature of the probe, i.e., the quantum particle. For this reason we stick with the nonrelativistic formulation. Relativistic effects solely amount to reducing the Hausdorff dimension to $D_H=1$ in the ultrarelativistic limit where the particle is confined to the lightcone \cite{relativistic}. As will be clear from the following discussion, this behavior due to relativistic effects does not counteract our conclusions. To this purpose, we recall that the nonrelativistic procedure has already been used in the context of the Hausdorff dimension of a quantum string, even if string excitations, which are crucial for calculating the Hausdorff dimension, occur in the ultrarelativistic regime \cite{Ansoldi:1997cw}. 
This line of reasoning is in analogy to what happens for the spectral dimension, since the diffusion equation is formally equivalent to a Wick rotated Schr\"{o}dinger equation for a nonrelativistic particle.

The crucial quantity is $\langle \Delta l \rangle $, i.e., the expectation value for traveled path length in time lapse $\Delta t$. To determine the new value for  (\ref{Deltat}) we need to calculate
\begin{equation}
\langle \textbf{y}|\hat{U}(\Delta t)|\psi\rangle  =\int \frac{d^d k}{(2\pi)^{\frac{d}{2}}} e^{-\frac{\ell^2 \textbf{k}^2}{(\Delta x)^2}} \langle \textbf{k} | \psi \rangle \ e^{i\textbf{ky}}e^{\frac{i\hbar\Delta t \textbf{k}^2}{2m(\Delta x)^2 }}
\end{equation}
where $\langle \textbf{k}|\psi\rangle$ is chosen as in (\ref{kprofile}). 
From (\ref{length}) we obtain the length of the path:
\begin{eqnarray}
\langle l \rangle &\propto& \frac{\hbar T}{m\Delta x} \left(1+\frac{\ell^2}{(\Delta x)^2}\right)^{-\frac{d+1}{2}}\nonumber\\ &&\times
\sqrt{1+\left(1+\frac{\ell^2}{(\Delta x)^2}\right)^2 \frac{4m^2(\Delta x)^4}{\hbar^2(\Delta t)^2}}.
\label{newlength}
\end{eqnarray}

  \begin{figure}
   \includegraphics[height=4.5cm]{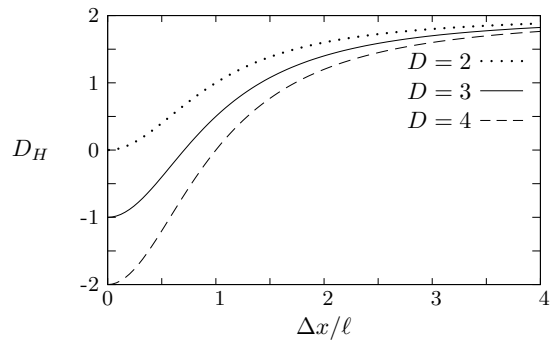}
  \hspace{0.2cm}
  		\caption{\label{Plot}The Hausdorff dimension of a particle path depending on $\Delta x/\ell$ for different numbers of spacetime dimensions.}
  		
   \end{figure}

As far as we consider length scales $\ell\ll \Delta x \ll \sqrt{\hbar \Delta t/m}$, the actual length of the path is an ill defined quantity which depends on the choice of the resolution parameter $\Delta x$, i.e., $\langle l \rangle\propto \hbar T/m\Delta x$, which matches the conventional result in (\ref{length2}). However (\ref{newlength}) presents an important new feature. The quantity $\langle l \rangle$ can never be infinite. Indeed in the limit $\Delta x\ll\ell$,  by keeping $\Delta t$ fixed one finds
\begin{equation}
\langle l \rangle\propto \frac{\hbar T}{m\ell}\left(\frac{\Delta x}{\ell}\right)^{d}.
\end{equation}
This is due to the presence of $\ell$, which nicely works as a natural cutoff in agreement with all the existing literature based on this formulation \cite{review1}. The problem is that $\ell$ is actually a minimal length, beyond which we lose the definition of position. In other words, for $\Delta x\ll\ell$ we are probing the microstructure of the manifold, which is affected by huge quantum geometry fluctuations. As a result the very concept of length is no longer meaningful, a fact which is confirmed by the vanishing value of $\langle l \rangle$.  Again we are left  with the only possibility of invoking the Hausdorff length to have some reliable information about the length of the path.  By using (\ref{hauslength}), we obtain
\begin{equation}
L_H \propto (\Delta x)^{D_H-2}\left(1+\frac{\ell^2}{(\Delta x)^2}\right)^{-\frac{d+1}{2}}.
\label{hauslength2}
\end{equation}

In the regime $\ell\ll \Delta x \ll \sqrt{\hbar \Delta t/m}$ we find the conventional result $D_H=2$. Conversely for $\Delta x\ll \ell$ we find $D_H=1-d$. This means that the Hausdorff dimension is either vanishing or negative. In fractal geometry this is the case of an \textit{empty set} \cite{negative}, which physically we could interpret as a ``dissolution'' of the path as far as trans-Planckian scales are probed. By requiring $\partial L_H/\partial (\Delta x)=0$ we can calculate the general form of the Hausdorff dimension which reads
\begin{equation}
 D_H=2 -\frac{d+1}{1+(\Delta x)^2/\ell^2}.
 \label{haus3}
 \end{equation}
 Some comments are in order. First, the Hausdorff dimension is always smaller than the usual value $2$. This is reminiscent of what we found when studying the spectral dimension in \cite{Modesto:2009qc}.  We recall that in a $D$-dimensional Euclidean geometry the heat equation reads
\begin{eqnarray}
\Delta K\left( x, y ; s \right)=
\frac{\partial}{\partial s}\,
K\left( x, y ; s \right)
\label{heat}
\end{eqnarray}
where $s$ is a fictitious diffusion time of dimension of a length squared, $\Delta$ is the Laplace operator, and $K\left( x ,y ; s \right)$ is the heat kernel, representing the probability density of diffusion from $x$ to $y$.
We showed that  the minimal length $\ell$, by introducing quantum gravity fuzziness, prevents the diffusion process to access to all the $D$ topological dimensions of the spacetime manifold. More specifically from (\ref{heat}) the spectral dimension for the flat space case turns out to be
\begin{eqnarray}
{\mathbb D}= \frac{s}{s+\ell^2}\ D.
\end{eqnarray}
i.e., it is smaller than the topological dimension of spacetime ${\mathbb D}<D=d+1$. In other words, while the uncertainty in quantum mechanics provides an erratic character to the path and a consequent increase of the Hausdorff dimension, the uncertainty in quantum gravity is responsible for resolution loss, whose amount is encoded in the difference $D-{\mathbb D}$. This fact becomes even more clear if we Wick rotate back the diffusion equation (\ref{heat}) and  we identify the diffusion time with $(\Delta x)^2$ by means of the relation (\ref{Deltatx}). As a result one finds
\begin{equation}
D_H=2- (D-{\mathbb D}).
\end{equation}
In the special case $D=2$, there is just one spatial dimension and the two indicators coincide: $D_H={\mathbb D}$.

  \begin{figure*}
   \includegraphics[width=17cm]{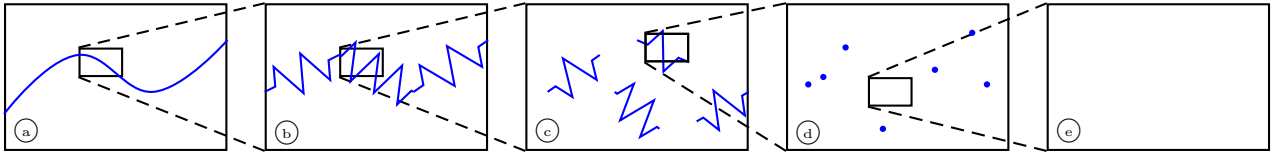}
  \hspace{0.2cm}
  		\caption{\label{magnification2}Schematic view of the geometrical structure of the quantum particle path in (a) the classical regime, (b) the quantum-mechanical regime, (c),(d) the Planckian regime, and (e) the trans-Planckian regime.}
  \end{figure*}

Second, one might ask whether the Hausdorff dimension assumes the classical value $D_H=1$. From (\ref{haus3}) this condition is met when $d=(\Delta x)^2/\ell^2$. As a result we have
\begin{equation}
\langle  L_H \rangle =\left\langle l\right\rangle \propto \frac{d^{d/2}}{(d+1)^{\frac{d+1}{2}}} \ \frac{\hbar T}{m\ell}
\end{equation}
which enjoys the desired feature of being independent of the resolution parameter $\Delta x$.
However we cannot interpret this result as a restoration of the classical character of the path. We should better say that we have another example in which $\ell$ provides a finite value for the proper length $\left\langle l\right\rangle$ of a fractal. Indeed the Hausdorff dimension can even descend below the value $D_H=1$ and beyond.
For scales $(d-1)/2=(\Delta x)^2/\ell^2$ we have $D_H=0$. The Hausdorff dimension reaches negative values corresponding to the case of empty sets.
The full behavior of $D_H$ can be seen in Fig. \ref{Plot}.

Third, there is the issue of self-similarity. The fact that the Hausdorff dimension actually can descend below the topological value $1$ is a sort of ``red flag.'' This happens when we are probing the path with resolution comparable with the size of the minimal length $\Delta x\sim\ell$. In other words, quantum gravity must introduce a length scale. At such a scale the manifold fluctuations are so strong that the path starts dissolving. As a result we have a breaking of the self-similarity or scale invariance property of the path. To check this result we just need to study the self-similarity condition $\langle \Delta l \rangle \propto \Delta x$. For $\ell\ll \Delta x \ll \sqrt{\hbar \Delta t/m}$ we just recover the conventional result as in (\ref{Deltatx}). Conversely for $\Delta x \lesssim \ell$ we get
\begin{equation}
\langle \Delta l \rangle\propto \ell \left(\frac{\Delta x}{\ell}\right)^{d+2}
\end{equation}
 which is different from (\ref{Deltatx}) unless one fixes $\Delta x=\ell$. This implies a breaking of scale invariance in the transition from one to the other regime. 

 We are now ready to draw conclusions. In reference to the three main goals we can say that
 \begin{enumerate}
\item the Hausdorff dimension we calculated in (\ref{haus3}) accounts for both the quantum mechanics uncertainty and the amount of fluctuations of the manifold through the term $\Delta x/\ell$. In addition $D_H$ depends also on the number of  dimensions $d$ of the manifold where the particle propagates;
\item the new feature we discovered through the study of $D_H$ is the expected scale invariance breaking as far as one introduces a length scale $\ell$ in the formalism;
\item the universal property we discovered in both indicators, i.e., $D_H$ and ${\mathbb D}$, is the amount of resolution loss $D-{\mathbb D}$ which affects both the path and the manifold when quantum gravity fluctuations occur.
\end{enumerate}
We could generalize our calculation to the case of nonvanishing average momentum $\textbf{p}_{\mathrm{av}}$. However one can prove that there is no additional physical information coming from it.
Essentially the conclusions we have just drawn are confirmed.

As a final point we can summarize our results for the character of a quantum path in the presence of a minimal length with the following scenario:  As long as $\Delta x\gg\sqrt{\hbar \Delta t/m}$ there exists a \textit{classical regime} in which the path is represented by a smooth differential curve, whose Hausdorff dimension coincides with the topological dimension of the curve, i.e., $D_H=1$; for $\ell\ll \Delta x\ll\sqrt{\hbar \Delta t/m}$ there is the \textit{quantum mechanics regime}, in which the path of the particle becomes strongly erratic and self-similar, approaching the configuration of a $D_H=2$ fractal; for smaller length scales, i.e.,
$\Delta x\sim\ell$, there is the \textit{Planckian regime}, which is characterized by a loss of resolution of the path with consequent decrease of the Hausdorff dimension and breaking of the self-similarity property; finally in the \textit{trans-Planckian regime}, i.e., $\Delta x\ll\ell$, the path is disintegrated by  huge fluctuations of the manifold and the Hausdorff dimension can be vanishing or even negative, corresponding to the case of an empty set (see Fig.\ref{magnification2}).

For the sake of truth we have to notice that, even if fascinating and mathematically correct, the interpretation of the trans-Planckian regime in terms of empty sets might be not fully physically consistent. We recall that our results are based on an effective approach to quantum gravity, which captures the feature of the emergence of a minimal length. We actually ignored the ultimate fate of a particle path as well as of a manifold in the trans-Planckian regime. Therefore we must keep our interpretation in terms of empty sets just as an indication for a possible scenario and assume reliability of our treatment only for length scales $\Delta x\gtrsim\ell$.
We also add that our results have been derived in terms of a specific approach to model the presence of a minimal length.  Yet our model captures some general feature of any effective approach to quantum gravity by squeezing momentum space integration measure at higher momenta. In other contexts, like black hole physics, this approach has led to model-independent descriptions \cite{NCBHs}. However we still miss in the context of the Hausdorff dimension a support to the universality of our results from other, hopefully full (i.e., noneffective), quantum gravity formulations. For this reason, we believe that the role of the Hausdorff dimension as an indicator of the fuzziness of a quantum manifold will deserve further investigations.



\begin{acknowledgments}
 P.N. is supported by the Helmholtz International Center for FAIR within the
framework of the LOEWE program (Landesoffensive zur Entwicklung Wissenschaftlich-\"{O}konomischer Exzellenz) launched by the State of Hesse. P.N. and B.N. thank the Perimeter Institute for Theoretical Physics, Waterloo, Ontario, Canada for the kind hospitality during the final period of work on this project.
\end{acknowledgments}


\end{document}